\documentclass[a4paper,twocolumn,superscriptaddress]{revtex4}
\usepackage[latin1]{inputenc}
\usepackage{graphicx}
\usepackage{amsfonts}
\usepackage{amsmath}
\usepackage{amssymb}
\usepackage{verbatim} 
\usepackage{latexsym}
\usepackage{indentfirst}
\usepackage{amsmath}
\usepackage{subfigure}
\usepackage{mathrsfs}

\begin{document}
\title{Effective Action for Hard Thermal Loops in Gravitational Fields}

\author{R. R. Francisco}
\affiliation{Universidade do Estado de Santa Catarina, 88330-668, Balneário Camboriú, SC, Brazil}
\author{J. Frenkel}
\affiliation{Instituto de Física, Universidade de São Paulo, 05508-090, São Paulo, SP, Brazil} 
\author{J. C. Taylor} \thanks{e-mail: rafael.francisco@udesc.br, jfrenkel@fmail.if.usp.br, \\ jct@damtp.cam.ac.uk}
\affiliation{DAMTP, Centre for Mathematical Sciences, Wilberforce Road, Cambridge, CB3 OWA, United Kingdom}

\begin{abstract}

We examine, through a Boltzmann equation approach, the generating action of hard thermal loops in the background of gravitational fields. Using the gauge and Weyl invariance of the theory at high temperature, we derive an explicit closed-form expression for the effective action.

\end{abstract}

\maketitle

\section{Introduction}

The study of thermal field theory in curved spacetimes is of interest due to the complex gauge structure of gravity and also because of its cosmological applications. In a curved spacetime which is asymptotically flat, one can consider matter as being in thermal equilibrium at temperature $T$, defined in the asymptotic Minkowski space. In such a spacetime, one can write the metric $g^{\mu\nu}$ in terms of the deviation from the Minkowski metric $\eta^{\mu\nu}$, for instance as

\begin{equation}
\label{eq1}
g^{\mu\nu}(x)=\eta^{\mu\nu}+\kappa h^{\mu\nu}(x),
\end{equation}
where $\kappa = \sqrt{32\pi G}$ and the graviton field $h^{\mu\nu}(x)$ tends to zero asymptotically.

\noindent{Making} an expansion in powers of $\kappa$, perturbation theory may be used to evaluate thermal loop diagrams in which the internal lines denote hot matter and the external lines represent the gravitational fields. However, in order to control the infrared divergences which occur in thermal perturbation theory, it is necessary first to calculate the hard thermal loops, where the external momenta are much smaller than the temperature $T$ \cite{1,2}.

There has been much work on thermal field theories in the presence of gravitational fields \cite{3,4,5} and general expressions for the partition functions, which generate the high-temperature amplitudes, were given in an implicit integral form \cite{6,7}. Moreover, in the special cases of static and long wavelength configurations of the gravitational fields, explicit expressions for the effective actions have also been presented \cite{5,8,9}.

The purpose of this work is to derive, to all orders in $\kappa$, an explicit and general expression for the effective action which generates the hard thermal loops in external gravitational fields. Our main result for the effective action is given in eq. \eqref{eq12}, which has the following properties:

\begin{itemize}
\item[a)] It is a non-local functional of the graviton fields, where the non-localities are line integrals in the direction of the hot particles momenta.
\item[b)] It is invariant under coordinate transformations that asymptotically reduce to the identity.
\item[c)] It is invariant under Weyl transformations 

\begin{equation}
\label{eq2}
g^{\mu\nu}(x)\rightarrow e^{2\sigma(x)}g^{\mu\nu}(x),
\end{equation}
where $\sigma(x)$ vanishes asymptotically.
\item[d)]It becomes, in the static and long wavelength limits, a local function of the external fields.
\end{itemize} 

The closed form \eqref{eq12}, which readily yields the results obtained from complicated Feynman diagrammatic computations \cite{10}, may be useful in a resummed perturbative calculation of thermal loops in the background of gravitational fields.

\section{The effective action}

Our derivation of the effective action $S(g)$ is based on the existence of a covariantly conserved energy-momentum tensor $T_{\mu\nu}$ such that:

\begin{equation}
\label{eq3}
\frac{\delta S(g)}{\delta g^{\mu\nu}(x)}=\frac{1}{2}\sqrt{-g}T_{\mu\nu}(x,g).
\end{equation}

This symmetric tensor can be related to a thermal distribution function $F(x,p,g)$ as

\begin{equation}
\label{eq4}
\sqrt{-g}T_{\mu\nu}(x,g)=\int d^{4}p\; p_{\mu}p_{\nu} F(x,p,g),
\end{equation}

where we have found it convenient to use the variables $x^{\alpha}$ and $p_{\beta}$, respectively, for the coordinates and momenta of the particles in a curved spacetime. 

In order to determine this distribution function, we use an extension of the method employed in \cite{7}, which relies upon a Boltzmann equation approach \cite{11,12}. Then, the Boltzmann equation for the distribution function $F(x,p,g)$ can be written in the simple form

\begin{equation}
\label{eq5}
(p\cdot\partial )F(x,p,g)= \hat{L}F(x,p,g),
\end{equation}
where $p\cdot\partial=\eta^{\mu\nu}p_{\nu}\partial/\partial x^{\mu}$ and $\hat{L}$ denotes a differential operator, which is linear in the graviton field $h^{\mu\nu}$ defined in \eqref{eq1}, given by:

\begin{equation}
\label{eq6}
\hat{L}=\frac{\kappa}{2}\bigg[\partial_{\lambda}h^{\mu\nu}(x)p_{\mu}p_{\nu}\frac{\partial}{\partial p_{\lambda}}-h^{\mu\nu}(x)(p_{\mu}\partial_{\nu}+p_{\nu}\partial_{\mu})\bigg].
\end{equation}

We seek a solution of the Boltzmann equation \eqref{eq5} which reduces to the equilibrium distribution at temperature T at infinity. It is now easy to find recursively such a solution, which has the form

\begin{equation}
\label{eq7}
F(x,p,g)=\sum_{n=0}^{\infty}\bigg(\frac{1}{p\cdot\partial}\hat{L}\bigg)^{n}F^{(0)}=\bigg(1-\frac{1}{p\cdot\partial}\hat{L}\bigg)^{-1}F^{(0)},
\end{equation}
where $F^{(0)}$ is the free distribution function:

\begin{equation}
\label{eq8}
F^{(0)}=\frac{2\mathscr{C}}{(2\pi)^{3}}\theta(p_{0})N(p_{0})\delta(\eta^{\alpha\beta}p_{\alpha}p_{\beta}-m^{2}).
\end{equation}

Here, $\mathscr{C}$ gives the number of internal degrees of freedom and

\begin{equation}
\label{eq9}
N(p_{0})=\frac{1}{e^{p_{0}/T}\pm 1}.
\end{equation}

In the following, the bare mass $m$ of the thermal particles will be neglected, which is a justified approximation at high temperature.

Inserting \eqref{eq7} and \eqref{eq4} in \eqref{eq3}, we see that the effective action $S(g)$ obeys the equation:

\begin{equation}
\label{eq10}
\frac{\delta S(g)}{\delta g^{\mu\nu}(x)}=\frac{1}{2}\int d^{4}p\; p_{\mu}p_{\nu}\bigg(1-\frac{1}{p\cdot\partial}\hat{L}\bigg)^{-1}F^{(0)}.
\end{equation}

In order to integrate this equation, we use the fact that the action $S(g)$ is a function of $\kappa h^{\mu\nu}$ (see \eqref{eq1}), which leads to the identity:

\begin{equation}
\label{eq11}
\kappa\frac{\delta S(g)}{\delta \kappa}=\int d^{4}x\; h^{\mu\nu}(x)\frac{\delta S(g)}{\delta h^{\mu\nu}(x)}.
\end{equation}

On the right hand side of this identity, we substitute $\delta S/\delta h^{\mu\nu}$ by $\kappa \delta S/ \delta g^{\mu\nu}$ as given by \eqref{eq10}. Then, integrating the ensuing relation over $\kappa$, we obtain the following expression for the effective action at high temperature

\begin{widetext}
\begin{equation}
\label{eq12}
S(g)=-\frac{\kappa}{2}\int d^{4}x\, h^{\mu\nu}(x)\int d^{4}p\;\bigg[ p_{\mu}p_{\nu}\hat{L}^{-1}(p\cdot\partial)\ln \bigg(1-\frac{1}{p\cdot\partial}\hat{L}\bigg)F^{(0)}\bigg],
\end{equation}
\end{widetext}
where $\hat{L}$ and $F^{(0)}$ are defined in (\ref{eq6}) and (\ref{eq8}), respectively. 

This expression represents an explicit closed-form result for the effective action which generates the hard thermal loops in external gravitational fields.

From \eqref{eq5}, which shows that $F(x, p, g)$ is a constant of motion in phase-space, it follows that the energy-momentum tensor \eqref{eq4} is covariantly conserved. Using this property in \eqref{eq3}, we deduce that the action $S(g)$ is invariant under coordinate transformations in an asymptotically minkowskian spacetime. Because of this invariance, the Ward identities require that the hard thermal loops with $n$ external lines have the same $T^{4}$ dependence for all $n$. 

Furthermore, since in the high temperature limit the particles are effectively massless, one may expect a Weyl invariance under the scale transformation \eqref{eq2}. Using \eqref{eq1} and \eqref{eq3}, this symmetry leads to the relation

\begin{equation}
\label{eq15}
(\eta^{\mu\nu}+\kappa h^{\mu\nu})\frac{\delta S}{\delta h^{\mu\nu}}=\frac{\kappa}{2}\sqrt{-g}g^{\mu\nu}T_{\mu\nu}=0,
\end{equation}
which accounts for the traceless property of the energy-momentum tensor at high temperature. We now verify that the equation \eqref{eq12} does have this Weyl invariance. To this end we note that (for $m=0$) the distribution function $F(x, p, g)$ which satisfies the Boltzmann equation \eqref{eq5}, can be written in the general form (compare with \eqref{eq8})

\begin{equation}
\label{eq16}
F(x, p, g)=2\mathcal{N}\mathscr{C}\theta(P_{0})N(P_{0})\delta[g^{\alpha\beta}(x)p_{\alpha}p_{\beta}],
\end{equation}
where $\mathcal{N}$ is a normalisation factor and $P_{0}$ is a constant of motion itself, which obeys \eqref{eq5}. Using this form, it is easy to see that the energy-momentum tensor $T_{\mu\nu}$ in \eqref{eq4} is indeed traceless.

The above features can be verified explicitly by expanding the action \eqref{eq12} in powers of $h$. This generates, in momentum space, a one-graviton tadpole

\begin{equation}
\label{eq17}
\Gamma_{\alpha\beta}=\frac{\widetilde{\delta S}}{\delta h^{\alpha\beta}}\bigg|_{h=0}=\frac{\mathscr{C}\pi^{2}\kappa T^{4}}{60}\int\frac{d\Omega}{4\pi}Q_{\alpha}Q_{\beta},
\end{equation}
where $Q_{\alpha}=(1, \hat{Q})$ and $\int d\Omega$ denotes an angular integration over the three-dimensional unit vector $\hat{Q}$.
 
Moreover, \eqref{eq12} generates a two-point graviton function given by:

\begin{widetext}
\begin{multline}
\label{eq18}
\Pi_{\alpha\beta, \mu\nu}(k)=\frac{\widetilde{\delta^{2}S}}{\delta h^{\alpha\beta}\delta h^{\mu\nu}}\bigg|_{h=0}=\frac{\mathscr{C}\pi^{2}\kappa^{2}}{120}T^{4}\int \frac{d\Omega}{4\pi}\bigg[k^{2}\frac{Q_{\alpha}Q_{\beta}Q_{\mu}Q_{\nu}}{(k\cdot Q)^{2}}-\\
\frac{k_{\alpha}Q_{\beta}Q_{\mu}Q_{\nu}+k_{\beta}Q_{\alpha}Q_{\mu}Q_{\nu}+k_{\mu}Q_{\alpha}Q_{\beta}Q_{\nu}+k_{\nu}Q_{\alpha}Q_{\beta}Q_{\mu}}{k\cdot Q} \bigg].
\end{multline}
\end{widetext}

These results are in agreement with those obtained in thermal perturbation theory from Feynman diagrammatic calculations (see Appendix). Using \eqref{eq17} and \eqref{eq18}, which have the same $T^{4}$ dependence, one can easily check the relation:

\begin{equation}
\label{eq19}
\eta^{\mu\nu}\Pi_{\alpha\beta, \mu\nu}+\kappa\Gamma_{\alpha\beta}=0,
\end{equation}
which is in accordance with the Weyl identity \eqref{eq15}.

The usual perturbative expansion of thermodynamic functions has in general a poor convergence, which stems from the fact that at high temperature the physical states are not described by massless particles. Instead, one must include plasma effects described by the hard thermal loops, which can be achieved through a reorganisation of the perturbative theory. In the resummed thermal perturbation theory one adds and subtracts the hard thermal loop effective action, which modifies the vertices and propagators in a self-consistent and gauge invariant manner \cite{13}. This procedure, which introduces thermal masses in the propagators, may control the infrared divergences and improve the convergence of the perturbative theory at finite temperature.

\section{The static and long wavelength limits}

Although the action \eqref{eq12} is, in general, a non-local functional of the external gravitational fields, it turns out that in the static and long wavelength limits, the effective action becomes a local function of the fields \cite{5, 7, 8, 9}. In order to understand this behaviour, we remark that the constant of motion $P_{0}$ which appears in the distribution function $F(x, p, g)$ in \eqref{eq16}, is given in the static and long wavelength limits respectively by:
\begin{subequations}
\label{eq20}
\begin{align}
P_{0}^{S}=&p_{0}; \\
P_{0}^{L}=&\sqrt{p_{i}p_{i}}.
\end{align}
\end{subequations}

This can be checked by noting that in these special cases, $p_{0}$ and $p_{i}$ satisfy the corresponding Boltzmann equation \eqref{eq5}. Then, it follows from \eqref{eq3}, \eqref{eq4} and \eqref{eq16} that, in these limits, the action $S^{S, L}(g)$ obeys the equation (where $\mathcal{N}^{S}= 1/(2\pi)^{3}$; $\mathcal{N}^{L}=1/[2(2\pi)^{3}]$)

\begin{multline}
\label{eq21}
\frac{\delta S^{S, L}}{\delta g^{\mu\nu}(x)}= \mathcal{N}^{S, L}\mathscr{C} \times \\
\int d^{4}p\; p_{\mu}p_{\nu}\theta(P_{0}^{S, L})N(P_{0}^{S,L})\delta[g^{\alpha\beta}(x)p_{\alpha}p_{\beta}].
\end{multline}

It can be verified that $P_{0}^{S}$ and $P_{0}^{L}$ are invariant respectively under time and space-independent coordinate transformations. Since these constants of motion are independent of the metric $g^{\mu\nu}$, \eqref{eq21} may be directly integrated. This leads to the result that, in the above limits, the effective action can be written in the local form:

\begin{widetext}
\begin{equation}
\label{eq22}
S^{S, L}(g)= \mathcal{N}^{S, L}\mathscr{C}\int d^{4}x \int d^{4}p\; \bigg\{\theta(P_{0}^{S, L})N(P_{0}^{S, L})\big\{\theta[g^{\alpha\beta}(x)p_{\alpha}p_{\beta}]-1\big\}\bigg\}.
\end{equation}
\end{widetext}

In addition of being gauge invariant in the static and long wavelength limits, respectively, the action \eqref{eq22} is also manifestly invariant under the Weyl transformation \eqref{eq2}. These properties and the symmetry of the hard thermal graviton amplitudes determine uniquelly the effective action. Thus, using the fact that $S^{S, L}$ agrees with $S$ to order $\kappa$, we deduce that \eqref{eq22} must agree to all orders with \eqref{eq12} evaluated in these limits. Indeed, performing the $p$-integrations in \eqref{eq22} one obtains \cite{7, 8, 14}, in the static and long wavelength limits, the correct local effective actions for hard thermal loops in an external gravitational field.
\\
\\

\noindent{\bf Acknowledgments}
\\

J. F. and R. R. F. would like to thank CNPq and FAPESP (Brazil) for financial support.
\appendix

\section*{APPENDIX}

We present here the result of a Feynman diagrammatic calculation which is in accordance with that given in \eqref{eq18}. We consider, for simplicity, the case of massless scalar particles in a gravitational field, which is described by the action

\begin{equation}
\label{a1} \tag{A1}
A=-\frac{1}{2}\int d^{4}x\; \sqrt{-g} g^{\mu\nu}(x)\partial_{\mu}\phi(x)\partial_{\nu}\phi(x),
\end{equation}
where the metric $g^{\mu\nu}$ is defined in \eqref{eq1}. Then, the corresponding contributions to the graviton self-energy come from the Feynman diagrams shown in figure 1.

\begin{figure}[!h]
\centering
\subfigure[]{\includegraphics[height=1.5cm]{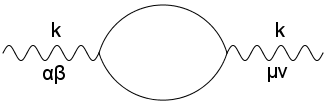}\;\;}
\subfigure[]{\includegraphics[height=2cm]{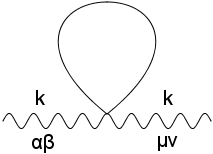}}
\caption{Lowest-order contributions to the thermal graviton two-point function.
Wavy lines represent gravitons and solid lines denote scalar particles.}
\end{figure}

\begin{widetext}
In the hard thermal loop approximation, graph (a) gives the contribution:

\begin{multline}
\label{a2} \tag{A2}
\Pi^{(a)}_{\alpha\beta, \mu\nu}(k)=\frac{\pi^{2}\kappa^{2}T^{4}}{120}\int \frac{d\Omega}{4\pi}\bigg[k^{2}\frac{Q_{\alpha}Q_{\beta}Q_{\mu}Q_{\nu}}{(k\cdot Q)^{2}}-\\
\frac{k_{\alpha}Q_{\beta}Q_{\mu}Q_{\nu}+k_{\beta}Q_{\alpha}Q_{\mu}Q_{\nu}+k_{\mu}Q_{\alpha}Q_{\beta}Q_{\nu}+k_{\nu}Q_{\alpha}Q_{\beta}Q_{\mu}}{k\cdot Q}+\eta_{\mu\nu}Q_{\alpha}Q_{\beta}+\eta_{\alpha\beta}Q_{\mu}Q_{\nu}\bigg].
\end{multline}

In a similar way, we obtain from graph (b) the contribution:

\begin{equation}
\label{a3} \tag{A3}
\Pi^{(b)}_{\alpha\beta, \mu\nu}(k)=-\frac{\pi^{2}\kappa^{2}T^{4}}{120}\int \frac{d\Omega}{4\pi}\bigg[\eta_{\mu\nu}Q_{\alpha}Q_{\beta}+\eta_{\alpha\beta}Q_{\mu}Q_{\nu}\bigg].
\end{equation}
\end{widetext}

Adding the results shown in \eqref{a2} and \eqref{a3} leads to the expression given in \eqref{eq18}, with $\mathscr{C}=1$, which is the appropriate weight factor for thermal scalar particles.

\end{document}